\documentclass[twocolumn,prl,aps,superscriptaddress,showpacs]{revtex4-1}   %

\usepackage{amsmath}
\usepackage{amssymb}
\usepackage{bbm}
\usepackage{braket}
\allowdisplaybreaks
\usepackage{pdfpages}

\usepackage{graphicx}
\usepackage[colorlinks=true,citecolor=blue,linkcolor=blue]{hyperref}

\newcommand{\equref}[1]{Eq.~(\ref{#1})}

\newcommand{\pdagger}{{\phantom{\dagger}}}
\newcommand{\bfj}{{\boldsymbol{j}}}

\newcommand{\up}{\uparrow}
\newcommand{\dw}{\downarrow}

\def\ie{\emph{i.e.},\ }

\renewcommand{\approx}{\simeq}

\renewcommand{\vec}[1]{\boldsymbol{#1}}

\begin{document}
\title{Dimensional crossover and cold-atom realization of topological Mott insulators}
\author{Mathias S. Scheurer}
\affiliation{Institute for Theory of Condensed Matter, Karlsruhe Institute of Technology (KIT), 76131 Karlsruhe, Germany}
\author{Stephan Rachel}
\affiliation{Institute for Theoretical Physics, TU Dresden, 01062 Dresden, Germany}
\author{Peter P.\ Orth}
\affiliation{Institute for Theory of Condensed Matter, Karlsruhe Institute of Technology (KIT), 76131 Karlsruhe, Germany}
\date{\today}
\begin{abstract}
We propose a cold-atom setup which allows for a dimensional crossover from a two-dimensional quantum spin Hall insulating phase to a three-dimensional strong topological insulator by tuning the hopping between the layers. We further show that additional Hubbard onsite interactions can give rise to spin liquid-like phases: weak and strong topological Mott insulators. They represent the celebrated paradigm of a quantum state of matter which merely exists because of the interplay of the non-trivial topology of the band structure and strong interactions. While the theoretical understanding of this phase has remained elusive, our proposal shall help to shed some light on this exotic state of matter by paving the way for a controlled experimental investigation in optical lattices.
\end{abstract}
\maketitle
Experiments of cold atoms in optical lattices allow to address fundamental questions of condensed matter physics~\cite{bloch-08rmp885,lewenstein-07ap243}. From a condensed matter perspective, those experiments accomplish two major achievements. First, they can serve as quantum simulators of systems that have a solid-state analogue with the crucial difference that one has a high degree of control and excellent tunability of most of the system parameters~\cite{Bloch2012}. One may even go beyond traditional models and study parameter regimes that are not accessible in the solid state.
On the other hand, cold-atom setups allow the realization of novel phases of matter and new phenomena that have been elusive in the solid state so far. Examples of the latter is the interaction driven Mott-superfluid quantum phase transition in the Bose-Hubbard model~\cite{greiner-02n39} or SU(N) magnetism using alkaline-earth atoms~\cite{Gorshkov-SUNColdAtoms-NatPhys-2010}.
Examples of the first case are investigations of Fermi-Hubbard models~\cite{hofstetter-02prl220407, esslinger-FermionicMI-Nature-2008, Schneider05122008, rey-09epl60001, lehur-09ap1452} in search for $d$-wave superfluidity or the study of three-dimensional Rashba spin-orbit coupling~\cite{PhysRevLett.108.235301}.

One of the most active areas of research in condensed matter physics at present are topological phases and, in particular, topological insulators~\cite{hasan-10rmp3045,qi-11rmp1057,bernevig2013}. These phases represent a new paradigm of quantum matter with a bulk gap but gapless surface states that are protected by symmetries such as time-reversal invariance. In real materials, the non-trivial band topology is usually caused by virtue of spin-orbit coupling. The realization of synthetic spin-orbit couplings for cold atoms~\cite{dalibard-11rmp1523,jaksch-03njp56, lin-11n83} thus marks the advent of topology to the field of ultra-cold gases. 

The dimensionality of space is an important parameter of topological phases in their classification according to anti-unitary symmetries~\cite{schnyder08}. The crossover between different dimensions is in general non-trivial: simply stacking the two-dimensional (2D) $\mathbb{Z}_2$ topological insulator~\cite{kane-05prl146802,kane-05prl226801,bernevig-06s1757,koenig-07s766} does not lead to its 3D counterpart, the strong topological insulator (STI)~\cite{moore-07prb121306,roy09prb195322,fu-07prl106803}. Instead, one obtains a 3D weak topological insulator (WTI)~\cite{hasan-11arcmp55}.

The situation becomes even more interesting in the presence of sufficiently strong interactions, where the topological classification\,\cite{schnyder08} breaks down and new phases can appear. As the interaction strength in cold-atom systems can be tuned over a wide range, they provide an ideal testbed to study the interplay of non-trivial topology and strong interactions. One of the most exciting theoretical proposals for such exotic states of matter is the {\it topological Mott insulator} of Pesin and Balents~\cite{pesin-10np376,footnote1}. It represents a spin liquid-like phase where the charge is stripped from the original electrons and frozen in a Mott insulating phase; the spinons (\ie the electrons' spin degree of freedom), however, still resemble the topological insulator and provide spin-only gapless surface states. The experimental observation of this state has so far remained elusive and even the theoretical understanding is rather limited~\cite{pesin-10np376,footnote1,rachel10,krempa-10prb165122,fiete11,cho-12njp114030}.

In this Letter, we propose an experimental setup which represents an example of both achievements of cold atoms in optical lattices: first, a quantum simulator for known phases which can be tuned into each other in a unique way.
Second, a realization of exotic  weak and strong topological Mott insulating phases which have not yet been found in solids. 

The model we propose is experimentally feasible and can be seen as a 3D generalization of an experimental setup for 2D topological insulators in optical lattices~\cite{goldman-10prl255302,cocks-12prl205303,orth-13jpb134004}. 
We show that the 2D system, which effectively consists of two time-reversed copies of massive Dirac Hamiltonians, can be tuned to a 3D weak or strong topological insulator just by varying a single hopping parameter. 
Since on-site Hubbard interactions are experimentally available, we analyze the interacting phase diagram within a slave-rotor theory~\cite{florens-02prb165111,florens-04prb035114, zhao-07prb195101}. Interactions drive a transition into a Mott insulating state, where the topological nature of the system survives and emerges as fractionalized gapless spin-only surface modes. Our proposal for a cold atom experiment and its possible subsequent realization will help to establish a deeper understanding  of this exotic phase.

The starting point of our analysis is the time-reversal invariant Hofstadter model on the square lattice~\cite{goldman-10prl255302}
\begin{align}
\begin{split}
 H_{2\text{D}} = -\sum_{\vec{j}}\Bigl(t\, &c_{\vec{j}+\vec{e}_x}^\dag e^{2\pi i\gamma\sigma_x} c^\pdagger_{\vec{j}} + t\,c_{\vec{j}+\vec{e}_y}^\dag e^{2\pi i\alpha j_x \sigma_z} c^\pdagger_{\vec{j}} \\ &+ \text{H.c.}\Bigr) + \lambda \sum_{\vec{j}} (-1)^{j_x} c_{\vec{j}}^\dag c^\pdagger_{\vec{j}} \,. \label{Nonint:Hamiltonian}\end{split}
\end{align}
It describes fermions with spin-1/2  in the presence of synthetic gauge fields; the parameter $\alpha$ describes the flux per plaquette of an artificial magnetic field perpendicular to the 2D plane, which due to the Pauli matrix $\sigma^z$ points in opposite directions for opposite spins. The spin mixing term $\gamma$ induces spin flips if the particle moves along the $x$-axis, and the $\lambda$-term describes a staggering of the optical lattice potential along the $x$-direction. All three terms can be experimentally realized~\cite{goldman-10prl255302}. 

\begin{figure}[tb]
\begin{center}
\includegraphics[width=\linewidth]{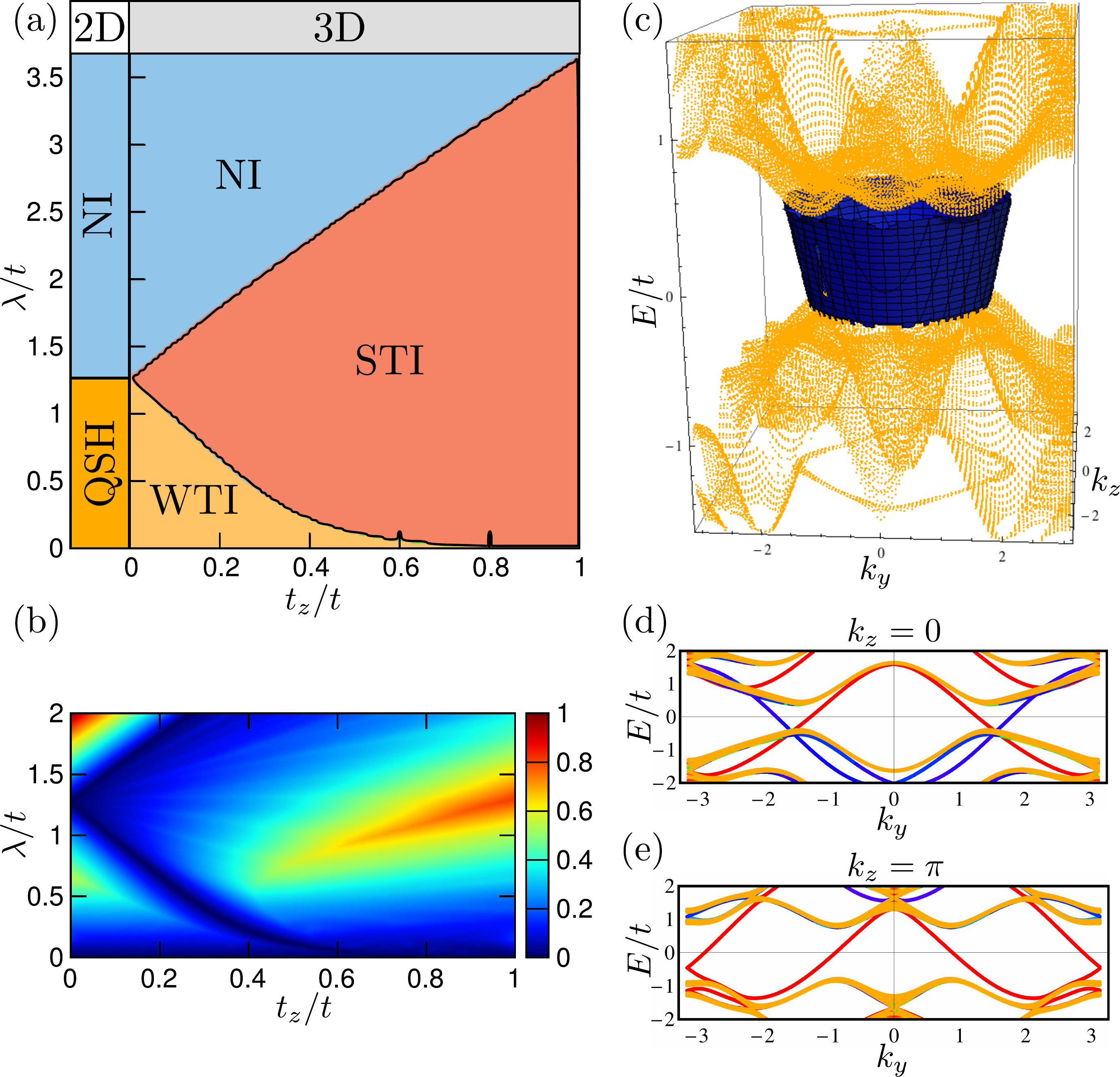}
\caption{(Color online) (a) 2D-3D crossover phase diagram as a function of layer coupling $t_z$ and staggered lattice potential $\lambda$. (b) Bulk gap as a function of $t_z$ and $\lambda$. (c) Surface state spectrum (blue) of the isotropic 3D system $t_x = t_y = t_z = t$, $\lambda=\lambda_c$ at the $x=0$ surface as a function of momenta $k_y$ and $k_z$. Bulk states are shown in yellow. We use open (periodic) boundary conditions along $x$ ($y, z$). (d-e) One-dimensional cuts of the spectrum for fixed values of $k_z = 0$ and $k_z = \pi$. Gapless edge states at $x=0$ ($x=L$) edge are shown in red (blue) yielding $z_0 = 1$ and $z_\pi = 0$ and thus $\nu_0 = z_0 + z_\pi= 1$.}
\label{NoninteractingPhases}
\end{center}
\end{figure}

The model belongs to the symmetry class AII \cite{schnyder08} and its phase diagram in 2D hosts quantum spin Hall (QSH) and normal insulating (NI) as well as (semi-) metallic phases~\cite{goldman-10prl255302, cocks-12prl205303,orth-13jpb134004}. The QSH phase is characterized by an odd number of helical edge states per edge, while the NI features an even number (including zero).
The topological $\mathbb{Z}_2$ invariant $\nu$ distinguishes between QSH ($\nu=1$) and NI ($\nu=0$) phase~\cite{kane-05prl226801}. It is sufficient to consider half filling and fixed values of $\alpha = 1/6$ and $\gamma = 1/4$, as the phase diagram as a function of the staggered potential $\lambda$ already contains the three phases. As shown in Fig.~\ref{NoninteractingPhases}(a) the 2D system is a QSH insulator for $|\lambda| < \lambda_c = 2^{1/3}t$ and becomes normal insulating for $|\lambda| > \lambda_c$~\cite{goldman-10prl255302, cocks-12prl205303,orth-13jpb134004}.
The system is semi-metallic at $\lambda = 0$ hosting six Dirac cones and at $\lambda = \lambda_c$
with one doubly degenerate Dirac cone.

We are interested in studying the dimensional crossover from two to three dimensions by continuously turning on a hopping parameter $t_z$ in the third direction that couples the different 2D layers. To obtain STI phases, we consider an interlayer coupling term of the form 
\begin{equation}
 H_z=-t_z\sum_{\vec{j}}\Bigl( \, c_{\vec{j}+\vec{e}_z}^\dag e^{2\pi i\alpha j_x \sigma_y} c^\pdagger_{\vec{j}} + \text{H.c.}\Bigr) \label{HoppingInZ} \,.
\end{equation} 
It contains a synthetic gauge field that represents an artificial magnetic field along the $y$-direction, which points in opposite directions for spins aligned parallel and anti-parallel to the $y$-axis. Note that this term respects time-reversal symmetry and, hence, the resulting 3D Hamiltonian $H_{2\text{D}}+H_z$ still belongs to class AII. 

As we have to deal with a 12-band model in 3D, some intuitive understanding is desirable. 
For this reason, we derive an effective theory valid in the vicinity of the point $(t_z,\lambda)=(0,\lambda_c)$ in the phase diagram of Fig.~\ref{NoninteractingPhases}(a), where all the distinct phases meet: QSH and NI (when $t_z=0$) as well as STI, WTI, and NI (when $t_z>0$). The doubly degenerate Dirac cone at this multi-critical point is formed out of four bands, the other eight bands are well-separated from the Fermi level.
Neglecting corrections quadratic in $t_z/t$, $(\lambda-\lambda_c)/t$ and $k_x, k_y$, quasi-degenerate perturbation theory \cite{WinklerSOC} yields the effective four-band Hamiltonian 
\begin{align}
 h^\text{eff}(\vec{k}) &= \begin{pmatrix} h_0(\vec{k}) & -ic\,\tau_xt_z \sin(k_z) \\ ic\,\tau_xt_z \sin(k_z) & h_0^*(-\vec{k}) \end{pmatrix}\,, 
\label{Nonint:EffHamiltonian}
 \end{align} 
where the upper left $2\times 2$ block is given by
\begin{align}
  h_0(\vec{k}) &= g_i(\vec{k})\tau_i ,\quad \vec{g}(\vec{k}) = \bigl(-at\, k_y, bt\, k_x, m(k_z)\bigr) \label{Nonint:EffHamiltonian2}
\end{align}
with mass $m(k_z)=d\,[(\lambda_c/t+\lambda_c^2/t^2) t_z \cos(k_z) - \delta \lambda]$ and $\delta \lambda = \lambda-\lambda_c$. The Pauli matrices $\tau_i$ act within the $2\times 2$-blocks in \equref{Nonint:EffHamiltonian} and $a$, $b$, $c$, $d$ are positive constants. For convenience, we have chosen the basis functions such that time-reversal is given by $is_y\mathcal{K}$, where the Pauli matrices $s_i$ act between the different $2\times 2$-blocks and $\mathcal{K}$ denotes complex conjugation. In addition, this convention directly reveals the connection of the 2D system ($t_z=0$) to the Bernevig-Hughes-Zhang-model \cite{bernevig-06s1757} and, consequently, we immediately know that $\delta \lambda=0$ marks the boundary between QSH and NI phases. As the regime $\delta \lambda >0$ is adiabatically connected to the topologically trivial limit $\lambda \rightarrow \infty$, the effective model reproduces the 2D phase diagram.

To continue with the 3D system, let us first emphasize that the effective Hamiltonian \eqref{Nonint:EffHamiltonian} is valid for the entire range  $-\pi < k_z \leq \pi$, since $t_z$ (and not $k_z$) has been taken as expansion parameter. In 3D, insulating phases are characterized by four $\mathbbm{Z}_2$ invariants $\nu_i$, with $i=0,1,2,3$, defined by invariants of 2D cuts of the 3D Brillouin zone~\cite{moore-07prb121306,roy09prb195322,fu-07prl106803}. To determine the strong invariant $\nu_0\equiv (z_0+z_\pi)\,\text{mod}\,2$, we need to calculate the invariants $z_0$ and $z_\pi$ associated with the time-reversal invariant planes $k_z=0$ and $k_z=\pi$. As \equref{Nonint:EffHamiltonian} conserves $s_z$ at $k_z=0,\pi$, spin-Chern numbers can be used~\cite{bernevig-06s1757,hasan-10rmp3045}.
One readily finds that if $m(0)$ and $m(\pi)$ have opposite signs, this will also hold for the Chern numbers and hence $\nu_0\equiv (z_0+z_\pi)\,\text{mod}\,2=1$. Therefore, the system is in the STI phase if and only if
\begin{equation}
 |\delta \lambda| < (\lambda_c/t + \lambda_c^2/t^2) t_z \simeq 2.85\,t_z.
\end{equation} 
Being adiabatically connected to $t_z = 0$, the other two phases ($|\delta \lambda| > 2.85\,t_z$) can easily be identified from the knowledge about the two-dimensional system. For $\delta \lambda > 2.85\,t_z$, we find a NI, whereas, in case of $\delta \lambda < - 2.85\,t_z$, the system resides in a WTI phase characterized by $(\nu_0;\nu_1,\nu_2,\nu_3)=(0;0,0,1)$~\cite{moore-07prb121306,roy09prb195322,fu-07prl106803}.

Finally, the effective Hamiltonian also allows to understand why a spin- and position-independent hopping term along the $z$-direction cannot result in an STI phase. Such a term would simply lead to a contribution proportional to the identity matrix in the $12$-band Bloch-Hamiltonian and thus to a term $\mathbbm{1}_{4\times 4} f(k_z)$ in the effective low-energy theory. Consequently, $\vec{g}$ in \equref{Nonint:EffHamiltonian2} would be independent of $k_z$ and hence $z_0=z_\pi$, excluding the appearance of an STI.

We verified our analysis of the effective model by numerically computing the $\mathbb{Z}_2$ invariant in all insulating phases of the full 12-band model using the approach of Ref.~\onlinecite{JPSJ.76.053702}. The corresponding phase diagram is illustrated in Fig.~\ref{NoninteractingPhases}(a) and is in perfect agreement with our previous discussion. The bulk gap is shown in Fig.~\ref{NoninteractingPhases}(b) and reaches its maximal value of the order of $t$ for isotropic hopping. Computing the spectrum with open boundary conditions along the $x$-direction, we clearly observe in Fig.~\ref{NoninteractingPhases}(c) (shown for $t_z=t$) a single gapless surface state (in blue) crossing the bulk gap (bulk bands in yellow). One-dimensional cuts of the spectrum are shown in Fig.~\ref{NoninteractingPhases}(d-e). The Dirac point of the surface cone is located inside the bulk bands (which is not unusual).

\begin{figure}[t!]
\begin{center}
\includegraphics[width=0.85\linewidth]{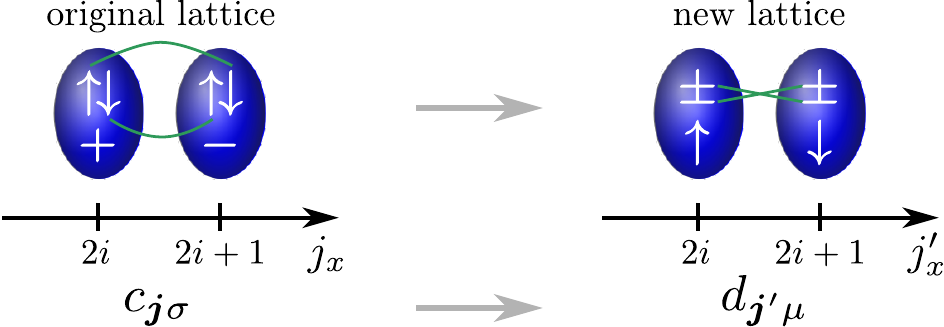}
\caption{(Color online) Twisting the Hubbard interaction via relabeling of spin and spatial even/odd sites along the $x$-direction. To obtain an identical Hamiltonian $H_{\text{2D}} + H_z$ one needs to experimentally implement different hopping elements (green lines) between nearest and next-nearest neighbor sites. For clarity, here we have only shown nearest neighbor hoppings. Details can be found in the Supplemental Material.}
\label{DefinitionOfTheMapping}
\end{center}
\end{figure}

While non-interacting topological phases are characterized by properties of their Bloch functions and can be classified according to symmetries, it is a largely open question what occurs in the presence of interactions. So far, mainly the effect of Hubbard onsite interactions $H_U=U \sum_{\bfj} n_{\bfj\uparrow} n_{\bfj\downarrow}$, where $n_{\bfj\sigma} = c_{\bfj\sigma}^\dag c_{\bfj\sigma}^{\phantom{\dagger}}$,  has been considered in order to address interaction effects in topological band structures. While this is very natural for solid state systems, the achievement of optical lattices allows to consider interactions which cannot be realized in real materials, opening a path towards exotic physics and giving rise to an even richer phenomenology. 
Here the key idea is to encode the spin degree of freedom $\sigma=\uparrow, \downarrow $ spatially and use the internal atomic hyperfine degree of freedom to represent the even/odd site information along $x$: $\mu = +$ for $j_x = 2 n$ and $\mu = -$ for $j_x = 2n+1$ with integer $n$. We denote the fermionic operators of the new lattice by $d_{\vec{j}'\mu}$ (see Fig.~\ref{DefinitionOfTheMapping}). Details of the associated mapping between the $c_{\vec{j}\sigma}$ and $d_{\vec{j}'\mu}$ fermions are given in the Supplementary Material. 

To keep the same non-interacting Hamiltonian $H_{\text{2D}} + H_{z}$ as before, one needs to experimentally implement different laser induced hopping elements for the $d_{\bfj' \mu}$ fermions. Only nearest neighbor and next-nearest neighbor terms along the three spatial directions are required.
As one can see in Fig.~\ref{DefinitionOfTheMapping}, in the new lattice, Hubbard onsite interactions do not couple $\up$-spin and $\dw$-spin on a given site but instead pairs of neighboring sites having the same spin orientation; the Hubbard term is twisted. The local Hubbard interaction on the new lattice reads in terms of the occupation numbers of the original $c_{\vec{j}\sigma}$ fermions as
\begin{align}
  \label{eq:2}
  H'_{U} &= \frac{U}{2} \sum_{\bfj, \sigma = \uparrow, \downarrow} (n_{2j_x j_y j_z\sigma} + n_{(2 j_x + 1) j_y j_z \sigma} - 1)^2\, .
\end{align}
It will generate topological Mott insulating (TMI) phases~\cite{pesin-10np376,footnote1} as shown below.
The TMI is characterized by a fractionalization of the original atoms into an internal and a number degree of freedom. In the TMI phase the atoms are localized, but their internal degree of freedom remains deconfined and inherits the non-trivial band topology of the original fermions. This state has been proposed to occur in Ir-based pyrochlore materials~\cite{pesin-10np376}, but has so far never been experimentally observed.

\begin{figure}[t!]
\begin{center}
\includegraphics[width=\linewidth]{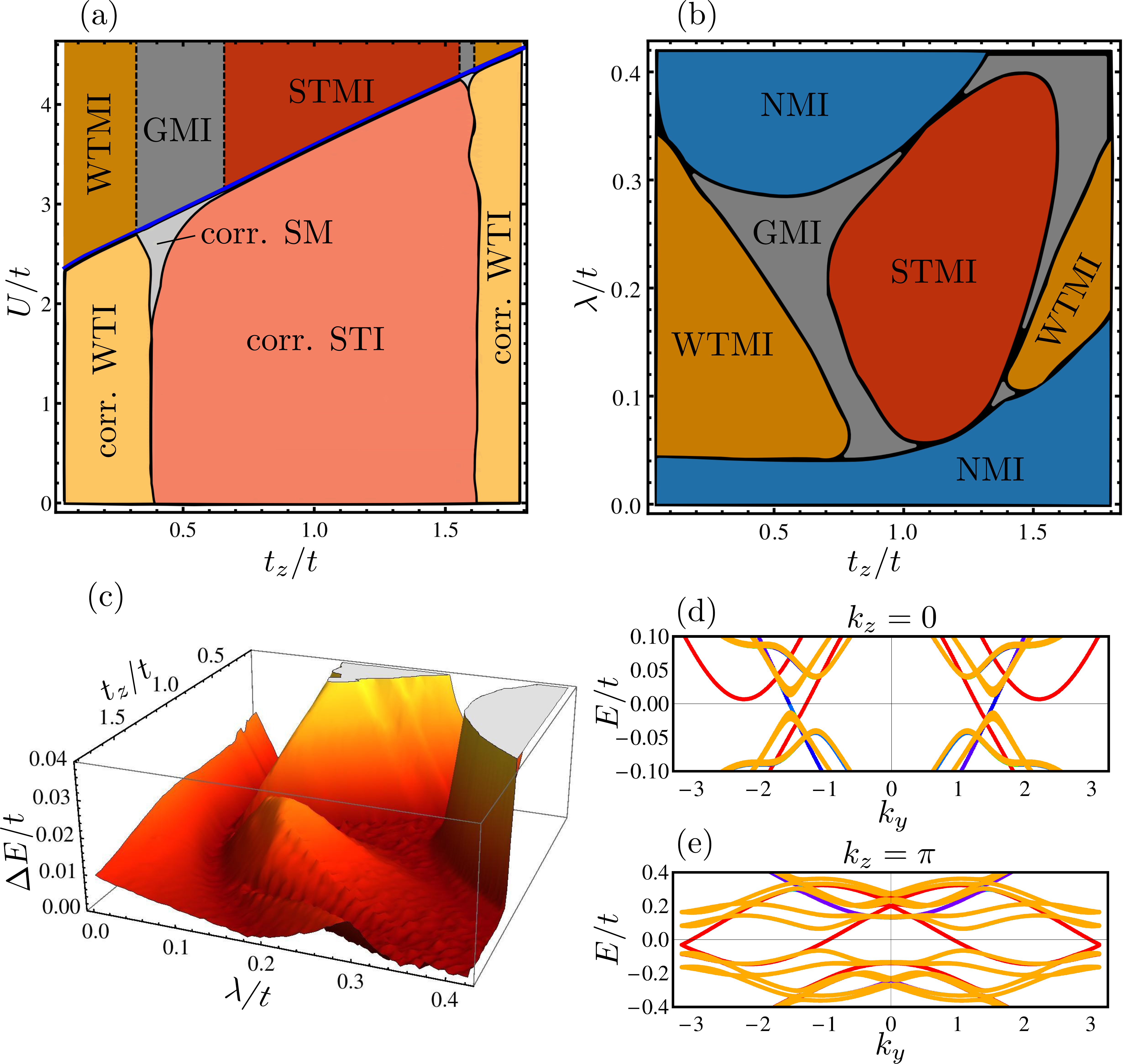}
\caption{(Color online) (a) Interacting phase diagram as a function of interaction $U$ and hopping $t_z$ for fixed $\lambda/t = 0.25$. Upon increasing $U$ the phases found at $U=0$ remain mostly intact with renormalized parameters. For $t_z/t \simeq 0.4$ an extended correlated semi-metallic (SM) phase appears at $U/t\simeq 2$. At a critical value $U_c/t \approx 2-4$ the system enters a Mott insulating (MI) state with weak and strong topological MI (WTMI and STMI) as well as gapless MI (GMI). The GMI phase exhibits a semi-metallic spinon spectrum. (b) Interacting phase diagram and (c) bulk gap at $U_c$ as a function of staggering $\lambda$ and hopping $t_z$. The STMI phase occupies a large part of the phase diagram and features a bulk gap as large as 10\% of the renormalized hopping parameters. (d-e) One-dimensional cut through the spinon bandstructure in the STMI phase for the isotropic system $t_x=t_y=t_z = t$ and $\lambda/t = 0.15$. We use open (periodic) boundary conditions along $x$ ($y,z$). Bulk states are shown in yellow, gapless spinon edge states at $x=0$ ($x=L$) edge are shown in red (blue).}
\label{InteractingPhases}
\end{center}
\end{figure}

We approach the TMI using slave-rotor theory \cite{florens-02prb165111,florens-04prb035114, zhao-07prb195101} which starts by writing the fermion operator as a product of number and internal degree of freedom $d_{\bfj \mu} = e^{i \theta_{\bfj}} f_{\bfj \mu}$. Here, $\theta_\bfj$ denote phases conjugate to the total particle number (of $d_{\bfj, \mu}$ fermions) on site $\bfj$ and $f_{\bfj \mu}$ is a spinon fermion operator that carries the internal index. The system is then described by two coupled mean-field Hamiltonians: a 3D quantum XY rotor model, which captures the number degrees of freedom and a renormalized non-interacting spinon Hamiltonian (see Suppl. Mat.). As the strength of the quantum fluctuations in the rotor model is determined by the interaction $U$, the rotor undergoes a transition from a ferro- to a paramagnetic state as $U$ is increased beyond a critical value $U_c$. This transition corresponds to the Mott transition. The spinons, on the other hand, are characterized by a bandstructure with 
renormalized parameters (set by the correlations between rotors), that can be topologically non-trivial and carry gapless surface excitations. 

Analyzing the interaction term\,\eqref{eq:2} within slave-rotor mean-field theory, we find that the interaction only renormalizes the hopping elementes $t_x, t_y, t_z$ but leaves $\lambda$ unchanged. It further induces slight spatial anisotropies between the hopping elements as well as inhomogeneities in the six atomic unit cell. Most importantly, the resulting spinon bandstructure can remain topological across the Mott transition. The resulting phase diagram, shown in Fig.~\ref{InteractingPhases}(a-b), thus exhibits the sought after strong and weak topological Mott insulator (STMI and WTMI) phases. These phases exhibit a bulk gap of the order of $10\%$ of the renormalized bandwidth (see Fig.~\ref{InteractingPhases}(c)). Due to the topological nature of the spinon bandstructure, they feature gapless spinon surface states shown in Fig.~\ref{InteractingPhases}(d-e) which are the defining property of the TMI phase.

Finally we discuss how the TMI phases are reflected in observable quantities. The atoms are frozen in a Mott insulating state which can be detected via standard time-of-flight measurements. The most straightforward way to detect the spinon surface states is to measure the spin-dependent spectral function or spin-current densities. Alternatively, one might also consider transport or thermodynamic quantities~\cite{Zhang02032012,arXiv:1306.4018}. For instance, the thermal conductivity shows a surface contribution which is expected to be linear in temperature~\cite{pesin-10np376}. Distinguishing between  WTMI and  STMI might be possible as they feature a different number of surface states. Spin-currents or quantities such as conductivity will therefore be different in the two phases.

To conclude, we have presented an experimentally feasible tight-binding Hamiltonian which allows to perform a quantum simulation of a dimensional crossover from a 2D quantum spin Hall to 3D strong and weak topological insulator phases. Considering the effect of on-site interactions, we have shown that weak and strong toplogical Mott insulator phases can be realized. Our results will hopefully lead to a subsequent experimental realization of this exotic state of matter.
\begin{acknowledgments}
The authors are grateful to Karyn Le Hur for early discussions on the project and acknowledge Walter Hofstetter, Daniel Cocks, Michael Buchhold, and Karyn Le Hur for previous collaborations on related topics. The Young Investigator Group of P.P.O. received financial support from the ``Concept for the Future'' of the KIT within the framework of the German Excellence Initiative. SR is supported by the DFG through FOR 960, the DFG priority program SPP 1666 ``Topological Insulators'', and by the Helmholtz association through VI-521.
\end{acknowledgments}


\begin{thebibliography}{42}%
\makeatletter
\providecommand \@ifxundefined [1]{%
 \@ifx{#1\undefined}
}%
\providecommand \@ifnum [1]{%
 \ifnum #1\expandafter \@firstoftwo
 \else \expandafter \@secondoftwo
 \fi
}%
\providecommand \@ifx [1]{%
 \ifx #1\expandafter \@firstoftwo
 \else \expandafter \@secondoftwo
 \fi
}%
\providecommand \natexlab [1]{#1}%
\providecommand \enquote  [1]{``#1''}%
\providecommand \bibnamefont  [1]{#1}%
\providecommand \bibfnamefont [1]{#1}%
\providecommand \citenamefont [1]{#1}%
\providecommand \href@noop [0]{\@secondoftwo}%
\providecommand \href [0]{\begingroup \@sanitize@url \@href}%
\providecommand \@href[1]{\@@startlink{#1}\@@href}%
\providecommand \@@href[1]{\endgroup#1\@@endlink}%
\providecommand \@sanitize@url [0]{\catcode `\\12\catcode `\$12\catcode
  `\&12\catcode `\#12\catcode `\^12\catcode `\_12\catcode `\%12\relax}%
\providecommand \@@startlink[1]{}%
\providecommand \@@endlink[0]{}%
\providecommand \url  [0]{\begingroup\@sanitize@url \@url }%
\providecommand \@url [1]{\endgroup\@href {#1}{\urlprefix }}%
\providecommand \urlprefix  [0]{URL }%
\providecommand \Eprint [0]{\href }%
\providecommand \doibase [0]{http://dx.doi.org/}%
\providecommand \selectlanguage [0]{\@gobble}%
\providecommand \bibinfo  [0]{\@secondoftwo}%
\providecommand \bibfield  [0]{\@secondoftwo}%
\providecommand \translation [1]{[#1]}%
\providecommand \BibitemOpen [0]{}%
\providecommand \bibitemStop [0]{}%
\providecommand \bibitemNoStop [0]{.\EOS\space}%
\providecommand \EOS [0]{\spacefactor3000\relax}%
\providecommand \BibitemShut  [1]{\csname bibitem#1\endcsname}%
\let\auto@bib@innerbib\@empty
\bibitem [{\citenamefont {Bloch}\ \emph {et~al.}(2008)\citenamefont {Bloch},
  \citenamefont {Dalibard},\ and\ \citenamefont {Zwerger}}]{bloch-08rmp885}%
  \BibitemOpen
  \bibfield  {author} {\bibinfo {author} {\bibfnamefont {I.}~\bibnamefont
  {Bloch}}, \bibinfo {author} {\bibfnamefont {J.}~\bibnamefont {Dalibard}}, \
  and\ \bibinfo {author} {\bibfnamefont {W.}~\bibnamefont {Zwerger}},\
  }\href@noop {} {\bibfield  {journal} {\bibinfo  {journal} {Rev. Mod. Phys.}\
  }\textbf {\bibinfo {volume} {80}},\ \bibinfo {pages} {885} (\bibinfo {year}
  {2008})}\BibitemShut {NoStop}%
\bibitem [{\citenamefont {{L}ewenstein}\ \emph {et~al.}(2007)\citenamefont
  {{L}ewenstein}, \citenamefont {{S}anpera}, \citenamefont {{A}hufinger},
  \citenamefont {{D}amski}, \citenamefont {{D}e},\ and\ \citenamefont
  {{S}en}}]{lewenstein-07ap243}%
  \BibitemOpen
  \bibfield  {author} {\bibinfo {author} {\bibfnamefont {M.}~\bibnamefont
  {{L}ewenstein}}, \bibinfo {author} {\bibfnamefont {A.}~\bibnamefont
  {{S}anpera}}, \bibinfo {author} {\bibfnamefont {V.}~\bibnamefont
  {{A}hufinger}}, \bibinfo {author} {\bibfnamefont {B.}~\bibnamefont
  {{D}amski}}, \bibinfo {author} {\bibfnamefont {A.~S.}\ \bibnamefont {{D}e}},
  \ and\ \bibinfo {author} {\bibfnamefont {U.}~\bibnamefont {{S}en}},\
  }\href@noop {} {\bibfield  {journal} {\bibinfo  {journal} {Adv. Phys.}\
  }\textbf {\bibinfo {volume} {56}},\ \bibinfo {pages} {243} (\bibinfo {year}
  {2007})}\BibitemShut {NoStop}%
\bibitem [{\citenamefont {Bloch}\ \emph {et~al.}(2012)\citenamefont {Bloch},
  \citenamefont {Dalibard},\ and\ \citenamefont {Nascimb\`{e}ne}}]{Bloch2012}%
  \BibitemOpen
  \bibfield  {author} {\bibinfo {author} {\bibfnamefont {I.}~\bibnamefont
  {Bloch}}, \bibinfo {author} {\bibfnamefont {J.}~\bibnamefont {Dalibard}}, \
  and\ \bibinfo {author} {\bibfnamefont {S.}~\bibnamefont {Nascimb\`{e}ne}},\
  }\href@noop {} {\bibfield  {journal} {\bibinfo  {journal} {Nat. Phys.}\
  }\textbf {\bibinfo {volume} {8}},\ \bibinfo {pages} {267} (\bibinfo {year}
  {2012})}\BibitemShut {NoStop}%
\bibitem [{\citenamefont {{G}reiner}\ \emph {et~al.}(2002)\citenamefont
  {{G}reiner}, \citenamefont {{M}andel}, \citenamefont {{E}sslinger},
  \citenamefont {{H}\"ansch},\ and\ \citenamefont {{B}loch}}]{greiner-02n39}%
  \BibitemOpen
  \bibfield  {author} {\bibinfo {author} {\bibfnamefont {M.}~\bibnamefont
  {{G}reiner}}, \bibinfo {author} {\bibfnamefont {O.}~\bibnamefont {{M}andel}},
  \bibinfo {author} {\bibfnamefont {T.}~\bibnamefont {{E}sslinger}}, \bibinfo
  {author} {\bibfnamefont {T.~W.}\ \bibnamefont {{H}\"ansch}}, \ and\ \bibinfo
  {author} {\bibfnamefont {I.}~\bibnamefont {{B}loch}},\ }\href@noop {}
  {\bibfield  {journal} {\bibinfo  {journal} {Nature (London)}\ }\textbf
  {\bibinfo {volume} {415}},\ \bibinfo {pages} {39} (\bibinfo {year}
  {2002})}\BibitemShut {NoStop}%
\bibitem [{\citenamefont {Gorshkov}\ \emph {et~al.}(2010)\citenamefont
  {Gorshkov}, \citenamefont {Hermele}, \citenamefont {Gurarie}, \citenamefont
  {Xu}, \citenamefont {Julienne}, \citenamefont {Ye}, \citenamefont {Zoller},
  \citenamefont {Demler}, \citenamefont {Lukin},\ and\ \citenamefont
  {Rey}}]{Gorshkov-SUNColdAtoms-NatPhys-2010}%
  \BibitemOpen
  \bibfield  {author} {\bibinfo {author} {\bibfnamefont {A.~V.}\ \bibnamefont
  {Gorshkov}}, \bibinfo {author} {\bibfnamefont {M.}~\bibnamefont {Hermele}},
  \bibinfo {author} {\bibfnamefont {V.}~\bibnamefont {Gurarie}}, \bibinfo
  {author} {\bibfnamefont {C.}~\bibnamefont {Xu}}, \bibinfo {author}
  {\bibfnamefont {P.~S.}\ \bibnamefont {Julienne}}, \bibinfo {author}
  {\bibfnamefont {J.}~\bibnamefont {Ye}}, \bibinfo {author} {\bibfnamefont
  {P.}~\bibnamefont {Zoller}}, \bibinfo {author} {\bibfnamefont
  {E.}~\bibnamefont {Demler}}, \bibinfo {author} {\bibfnamefont {M.~D.}\
  \bibnamefont {Lukin}}, \ and\ \bibinfo {author} {\bibfnamefont {A.~M.}\
  \bibnamefont {Rey}},\ }\href@noop {} {\bibfield  {journal} {\bibinfo
  {journal} {Nat. Phys.}\ }\textbf {\bibinfo {volume} {6}},\ \bibinfo {pages}
  {289} (\bibinfo {year} {2010})}\BibitemShut {NoStop}%
\bibitem [{\citenamefont {Hofstetter}\ \emph {et~al.}(2002)\citenamefont
  {Hofstetter}, \citenamefont {Cirac}, \citenamefont {Zoller}, \citenamefont
  {Demler},\ and\ \citenamefont {Lukin}}]{hofstetter-02prl220407}%
  \BibitemOpen
  \bibfield  {author} {\bibinfo {author} {\bibfnamefont {W.}~\bibnamefont
  {Hofstetter}}, \bibinfo {author} {\bibfnamefont {J.~I.}\ \bibnamefont
  {Cirac}}, \bibinfo {author} {\bibfnamefont {P.}~\bibnamefont {Zoller}},
  \bibinfo {author} {\bibfnamefont {E.}~\bibnamefont {Demler}}, \ and\ \bibinfo
  {author} {\bibfnamefont {M.~D.}\ \bibnamefont {Lukin}},\ }\href@noop {}
  {\bibfield  {journal} {\bibinfo  {journal} {Phys. Rev. Lett.}\ }\textbf
  {\bibinfo {volume} {89}},\ \bibinfo {pages} {220407} (\bibinfo {year}
  {2002})}\BibitemShut {NoStop}%
\bibitem [{\citenamefont {{J}\"ordens}\ \emph {et~al.}(2008)\citenamefont
  {{J}\"ordens}, \citenamefont {{S}trohmaier}, \citenamefont {{G}\"unter},
  \citenamefont {{M}oritz},\ and\ \citenamefont
  {{E}sslinger}}]{esslinger-FermionicMI-Nature-2008}%
  \BibitemOpen
  \bibfield  {author} {\bibinfo {author} {\bibfnamefont {R.}~\bibnamefont
  {{J}\"ordens}}, \bibinfo {author} {\bibfnamefont {N.}~\bibnamefont
  {{S}trohmaier}}, \bibinfo {author} {\bibfnamefont {K.}~\bibnamefont
  {{G}\"unter}}, \bibinfo {author} {\bibfnamefont {H.}~\bibnamefont
  {{M}oritz}}, \ and\ \bibinfo {author} {\bibfnamefont {T.}~\bibnamefont
  {{E}sslinger}},\ }\href@noop {} {\bibfield  {journal} {\bibinfo  {journal}
  {Nature (London)}\ }\textbf {\bibinfo {volume} {455}},\ \bibinfo {pages}
  {204} (\bibinfo {year} {2008})}\BibitemShut {NoStop}%
\bibitem [{\citenamefont {Schneider}\ \emph {et~al.}(2008)\citenamefont
  {Schneider}, \citenamefont {Hackermüller}, \citenamefont {Will},
  \citenamefont {Best}, \citenamefont {Bloch}, \citenamefont {Costi},
  \citenamefont {Helmes}, \citenamefont {Rasch},\ and\ \citenamefont
  {Rosch}}]{Schneider05122008}%
  \BibitemOpen
  \bibfield  {author} {\bibinfo {author} {\bibfnamefont {U.}~\bibnamefont
  {Schneider}}, \bibinfo {author} {\bibfnamefont {L.}~\bibnamefont
  {Hackermüller}}, \bibinfo {author} {\bibfnamefont {S.}~\bibnamefont {Will}},
  \bibinfo {author} {\bibfnamefont {T.}~\bibnamefont {Best}}, \bibinfo {author}
  {\bibfnamefont {I.}~\bibnamefont {Bloch}}, \bibinfo {author} {\bibfnamefont
  {T.~A.}\ \bibnamefont {Costi}}, \bibinfo {author} {\bibfnamefont {R.~W.}\
  \bibnamefont {Helmes}}, \bibinfo {author} {\bibfnamefont {D.}~\bibnamefont
  {Rasch}}, \ and\ \bibinfo {author} {\bibfnamefont {A.}~\bibnamefont
  {Rosch}},\ }\href@noop {} {\bibfield  {journal} {\bibinfo  {journal}
  {Science}\ }\textbf {\bibinfo {volume} {322}},\ \bibinfo {pages} {1520}
  (\bibinfo {year} {2008})}\BibitemShut {NoStop}%
\bibitem [{\citenamefont {{R}ey}\ \emph {et~al.}(2009)\citenamefont {{R}ey},
  \citenamefont {{S}ensarma}, \citenamefont {{F}\"olling}, \citenamefont
  {{G}reiner}, \citenamefont {{D}emler},\ and\ \citenamefont
  {{L}ukin}}]{rey-09epl60001}%
  \BibitemOpen
  \bibfield  {author} {\bibinfo {author} {\bibfnamefont {A.~M.}\ \bibnamefont
  {{R}ey}}, \bibinfo {author} {\bibfnamefont {R.}~\bibnamefont {{S}ensarma}},
  \bibinfo {author} {\bibfnamefont {S.}~\bibnamefont {{F}\"olling}}, \bibinfo
  {author} {\bibfnamefont {M.}~\bibnamefont {{G}reiner}}, \bibinfo {author}
  {\bibfnamefont {E.}~\bibnamefont {{D}emler}}, \ and\ \bibinfo {author}
  {\bibfnamefont {M.~D.}\ \bibnamefont {{L}ukin}},\ }\href@noop {} {\bibfield
  {journal} {\bibinfo  {journal} {EPL}\ }\textbf {\bibinfo {volume} {87}},\
  \bibinfo {pages} {60001} (\bibinfo {year} {2009})}\BibitemShut {NoStop}%
\bibitem [{\citenamefont {{L}e {H}ur}\ and\ \citenamefont
  {{R}ice}(2009)}]{lehur-09ap1452}%
  \BibitemOpen
  \bibfield  {author} {\bibinfo {author} {\bibfnamefont {K.}~\bibnamefont {{L}e
  {H}ur}}\ and\ \bibinfo {author} {\bibfnamefont {T.~M.}\ \bibnamefont
  {{R}ice}},\ }\href@noop {} {\bibfield  {journal} {\bibinfo  {journal} {Ann.
  Phys. (NY)}\ }\textbf {\bibinfo {volume} {324}},\ \bibinfo {pages} {1452}
  (\bibinfo {year} {2009})}\BibitemShut {NoStop}%
\bibitem [{\citenamefont {Anderson}\ \emph {et~al.}(2012)\citenamefont
  {Anderson}, \citenamefont {Juzeli\ifmmode~\bar{u}\else \={u}\fi{}nas},
  \citenamefont {Galitski},\ and\ \citenamefont
  {Spielman}}]{PhysRevLett.108.235301}%
  \BibitemOpen
  \bibfield  {author} {\bibinfo {author} {\bibfnamefont {B.~M.}\ \bibnamefont
  {Anderson}}, \bibinfo {author} {\bibfnamefont {G.}~\bibnamefont
  {Juzeli\ifmmode~\bar{u}\else \={u}\fi{}nas}}, \bibinfo {author}
  {\bibfnamefont {V.~M.}\ \bibnamefont {Galitski}}, \ and\ \bibinfo {author}
  {\bibfnamefont {I.~B.}\ \bibnamefont {Spielman}},\ }\href@noop {} {\bibfield
  {journal} {\bibinfo  {journal} {Phys. Rev. Lett.}\ }\textbf {\bibinfo
  {volume} {108}},\ \bibinfo {pages} {235301} (\bibinfo {year}
  {2012})}\BibitemShut {NoStop}%
\bibitem [{\citenamefont {Hasan}\ and\ \citenamefont
  {Kane}(2010)}]{hasan-10rmp3045}%
  \BibitemOpen
  \bibfield  {author} {\bibinfo {author} {\bibfnamefont {M.~Z.}\ \bibnamefont
  {Hasan}}\ and\ \bibinfo {author} {\bibfnamefont {C.~L.}\ \bibnamefont
  {Kane}},\ }\href@noop {} {\bibfield  {journal} {\bibinfo  {journal} {Rev.
  Mod. Phys.}\ }\textbf {\bibinfo {volume} {82}},\ \bibinfo {pages} {3045}
  (\bibinfo {year} {2010})}\BibitemShut {NoStop}%
\bibitem [{\citenamefont {Qi}\ and\ \citenamefont
  {Zhang}(2011)}]{qi-11rmp1057}%
  \BibitemOpen
  \bibfield  {author} {\bibinfo {author} {\bibfnamefont {X.-L.}\ \bibnamefont
  {Qi}}\ and\ \bibinfo {author} {\bibfnamefont {S.-C.}\ \bibnamefont {Zhang}},\
  }\href@noop {} {\bibfield  {journal} {\bibinfo  {journal} {Rev. Mod. Phys.}\
  }\textbf {\bibinfo {volume} {83}},\ \bibinfo {pages} {1057} (\bibinfo {year}
  {2011})}\BibitemShut {NoStop}%
\bibitem [{\citenamefont {Bernevig}(2013)}]{bernevig2013}%
  \BibitemOpen
  \bibfield  {author} {\bibinfo {author} {\bibfnamefont {B.~A.}\ \bibnamefont
  {Bernevig}},\ }\href@noop {} {\emph {\bibinfo {title} {Topological Insulators
  and Topological Superconductors}}}\ (\bibinfo  {publisher} {Princeton
  University Press},\ \bibinfo {address} {Princeton and Oxford},\ \bibinfo
  {year} {2013})\BibitemShut {NoStop}%
\bibitem [{\citenamefont {Dalibard}\ \emph {et~al.}(2011)\citenamefont
  {Dalibard}, \citenamefont {Gerbier}, \citenamefont
  {Juzeli\ifmmode~\bar{u}\else \={u}\fi{}nas},\ and\ \citenamefont
  {\"Ohberg}}]{dalibard-11rmp1523}%
  \BibitemOpen
  \bibfield  {author} {\bibinfo {author} {\bibfnamefont {J.}~\bibnamefont
  {Dalibard}}, \bibinfo {author} {\bibfnamefont {F.}~\bibnamefont {Gerbier}},
  \bibinfo {author} {\bibfnamefont {G.}~\bibnamefont
  {Juzeli\ifmmode~\bar{u}\else \={u}\fi{}nas}}, \ and\ \bibinfo {author}
  {\bibfnamefont {P.}~\bibnamefont {\"Ohberg}},\ }\href@noop {} {\bibfield
  {journal} {\bibinfo  {journal} {Rev. Mod. Phys.}\ }\textbf {\bibinfo {volume}
  {83}},\ \bibinfo {pages} {1523} (\bibinfo {year} {2011})}\BibitemShut
  {NoStop}%
\bibitem [{\citenamefont {Jaksch}\ and\ \citenamefont
  {Zoller}(2003)}]{jaksch-03njp56}%
  \BibitemOpen
  \bibfield  {author} {\bibinfo {author} {\bibfnamefont {D.}~\bibnamefont
  {Jaksch}}\ and\ \bibinfo {author} {\bibfnamefont {P.}~\bibnamefont
  {Zoller}},\ }\href@noop {} {\bibfield  {journal} {\bibinfo  {journal} {New.
  J. Phys.}\ }\textbf {\bibinfo {volume} {5}},\ \bibinfo {pages} {56} (\bibinfo
  {year} {2003})}\BibitemShut {NoStop}%
\bibitem [{\citenamefont {Lin}\ \emph {et~al.}(2011)\citenamefont {Lin},
  \citenamefont {Jimenez-Garcia},\ and\ \citenamefont {Spielman}}]{lin-11n83}%
  \BibitemOpen
  \bibfield  {author} {\bibinfo {author} {\bibfnamefont {Y.-J.}\ \bibnamefont
  {Lin}}, \bibinfo {author} {\bibfnamefont {K.}~\bibnamefont {Jimenez-Garcia}},
  \ and\ \bibinfo {author} {\bibfnamefont {I.~B.}\ \bibnamefont {Spielman}},\
  }\href@noop {} {\bibfield  {journal} {\bibinfo  {journal} {Nature}\ }\textbf
  {\bibinfo {volume} {471}},\ \bibinfo {pages} {83} (\bibinfo {year}
  {2011})}\BibitemShut {NoStop}%
\bibitem [{\citenamefont {Schnyder}\ \emph {et~al.}(2008)\citenamefont
  {Schnyder}, \citenamefont {Ryu}, \citenamefont {Furusaki},\ and\
  \citenamefont {Ludwig}}]{schnyder08}%
  \BibitemOpen
  \bibfield  {author} {\bibinfo {author} {\bibfnamefont {A.~P.}\ \bibnamefont
  {Schnyder}}, \bibinfo {author} {\bibfnamefont {S.}~\bibnamefont {Ryu}},
  \bibinfo {author} {\bibfnamefont {A.}~\bibnamefont {Furusaki}}, \ and\
  \bibinfo {author} {\bibfnamefont {A.~W.~W.}\ \bibnamefont {Ludwig}},\
  }\href@noop {} {\bibfield  {journal} {\bibinfo  {journal} {Phys. Rev. B}\
  }\textbf {\bibinfo {volume} {78}},\ \bibinfo {pages} {195125} (\bibinfo
  {year} {2008})}\BibitemShut {NoStop}%
\bibitem [{\citenamefont {Kane}\ and\ \citenamefont
  {Mele}(2005{\natexlab{a}})}]{kane-05prl146802}%
  \BibitemOpen
  \bibfield  {author} {\bibinfo {author} {\bibfnamefont {C.~L.}\ \bibnamefont
  {Kane}}\ and\ \bibinfo {author} {\bibfnamefont {E.~J.}\ \bibnamefont
  {Mele}},\ }\href@noop {} {\bibfield  {journal} {\bibinfo  {journal} {Phys.
  Rev. Lett.}\ }\textbf {\bibinfo {volume} {95}},\ \bibinfo {pages} {146802}
  (\bibinfo {year} {2005}{\natexlab{a}})}\BibitemShut {NoStop}%
\bibitem [{\citenamefont {Kane}\ and\ \citenamefont
  {Mele}(2005{\natexlab{b}})}]{kane-05prl226801}%
  \BibitemOpen
  \bibfield  {author} {\bibinfo {author} {\bibfnamefont {C.~L.}\ \bibnamefont
  {Kane}}\ and\ \bibinfo {author} {\bibfnamefont {E.~J.}\ \bibnamefont
  {Mele}},\ }\href@noop {} {\bibfield  {journal} {\bibinfo  {journal} {Phys.
  Rev. Lett.}\ }\textbf {\bibinfo {volume} {95}},\ \bibinfo {pages} {226801}
  (\bibinfo {year} {2005}{\natexlab{b}})}\BibitemShut {NoStop}%
\bibitem [{\citenamefont {Bernevig}\ \emph {et~al.}(2006)\citenamefont
  {Bernevig}, \citenamefont {Hughes},\ and\ \citenamefont
  {Zhang}}]{bernevig-06s1757}%
  \BibitemOpen
  \bibfield  {author} {\bibinfo {author} {\bibfnamefont {B.~A.}\ \bibnamefont
  {Bernevig}}, \bibinfo {author} {\bibfnamefont {T.~L.}\ \bibnamefont
  {Hughes}}, \ and\ \bibinfo {author} {\bibfnamefont {S.-C.}\ \bibnamefont
  {Zhang}},\ }\href@noop {} {\bibfield  {journal} {\bibinfo  {journal}
  {Science}\ }\textbf {\bibinfo {volume} {314}},\ \bibinfo {pages} {1757}
  (\bibinfo {year} {2006})}\BibitemShut {NoStop}%
\bibitem [{\citenamefont {K{\"o}nig}\ \emph {et~al.}(2007)\citenamefont
  {K{\"o}nig}, \citenamefont {Wiedmann}, \citenamefont {Br{\"u}ne},
  \citenamefont {Roth}, \citenamefont {Buhmann}, \citenamefont {Molenkamp},
  \citenamefont {Qi},\ and\ \citenamefont {Zhang}}]{koenig-07s766}%
  \BibitemOpen
  \bibfield  {author} {\bibinfo {author} {\bibfnamefont {M.}~\bibnamefont
  {K{\"o}nig}}, \bibinfo {author} {\bibfnamefont {S.}~\bibnamefont {Wiedmann}},
  \bibinfo {author} {\bibfnamefont {C.}~\bibnamefont {Br{\"u}ne}}, \bibinfo
  {author} {\bibfnamefont {A.}~\bibnamefont {Roth}}, \bibinfo {author}
  {\bibfnamefont {H.}~\bibnamefont {Buhmann}}, \bibinfo {author} {\bibfnamefont
  {L.~W.}\ \bibnamefont {Molenkamp}}, \bibinfo {author} {\bibfnamefont {X.-L.}\
  \bibnamefont {Qi}}, \ and\ \bibinfo {author} {\bibfnamefont {S.-C.}\
  \bibnamefont {Zhang}},\ }\href@noop {} {\bibfield  {journal} {\bibinfo
  {journal} {Science}\ }\textbf {\bibinfo {volume} {318}},\ \bibinfo {pages}
  {766} (\bibinfo {year} {2007})}\BibitemShut {NoStop}%
\bibitem [{\citenamefont {Moore}\ and\ \citenamefont
  {Balents}(2007)}]{moore-07prb121306}%
  \BibitemOpen
  \bibfield  {author} {\bibinfo {author} {\bibfnamefont {J.~E.}\ \bibnamefont
  {Moore}}\ and\ \bibinfo {author} {\bibfnamefont {L.}~\bibnamefont
  {Balents}},\ }\href@noop {} {\bibfield  {journal} {\bibinfo  {journal} {Phys.
  Rev. B}\ }\textbf {\bibinfo {volume} {75}},\ \bibinfo {pages} {121306(R)}
  (\bibinfo {year} {2007})}\BibitemShut {NoStop}%
\bibitem [{\citenamefont {Roy}(2009)}]{roy09prb195322}%
  \BibitemOpen
  \bibfield  {author} {\bibinfo {author} {\bibfnamefont {R.}~\bibnamefont
  {Roy}},\ }\href@noop {} {\bibfield  {journal} {\bibinfo  {journal} {Phys.
  Rev. B}\ }\textbf {\bibinfo {volume} {79}},\ \bibinfo {pages} {195322}
  (\bibinfo {year} {2009})}\BibitemShut {NoStop}%
\bibitem [{\citenamefont {Fu}\ \emph {et~al.}(2007)\citenamefont {Fu},
  \citenamefont {Kane},\ and\ \citenamefont {Mele}}]{fu-07prl106803}%
  \BibitemOpen
  \bibfield  {author} {\bibinfo {author} {\bibfnamefont {L.}~\bibnamefont
  {Fu}}, \bibinfo {author} {\bibfnamefont {C.~L.}\ \bibnamefont {Kane}}, \ and\
  \bibinfo {author} {\bibfnamefont {E.~J.}\ \bibnamefont {Mele}},\ }\href@noop
  {} {\bibfield  {journal} {\bibinfo  {journal} {Phys. Rev. Lett.}\ }\textbf
  {\bibinfo {volume} {98}},\ \bibinfo {pages} {106803} (\bibinfo {year}
  {2007})}\BibitemShut {NoStop}%
\bibitem [{\citenamefont {Hasan}\ and\ \citenamefont
  {Moore}(2011)}]{hasan-11arcmp55}%
  \BibitemOpen
  \bibfield  {author} {\bibinfo {author} {\bibfnamefont {M.~Z.}\ \bibnamefont
  {Hasan}}\ and\ \bibinfo {author} {\bibfnamefont {J.~E.}\ \bibnamefont
  {Moore}},\ }\href@noop {} {\bibfield  {journal} {\bibinfo  {journal} {Ann.
  Rev. Cond. Mat. Phys.}\ }\textbf {\bibinfo {volume} {2}},\ \bibinfo {pages}
  {55} (\bibinfo {year} {2011})}\BibitemShut {NoStop}%
\bibitem [{\citenamefont {Pesin}\ and\ \citenamefont
  {Balents}(2010)}]{pesin-10np376}%
  \BibitemOpen
  \bibfield  {author} {\bibinfo {author} {\bibfnamefont {D.~A.}\ \bibnamefont
  {Pesin}}\ and\ \bibinfo {author} {\bibfnamefont {L.}~\bibnamefont
  {Balents}},\ }\href@noop {} {\bibfield  {journal} {\bibinfo  {journal}
  {Nature Phys.}\ }\textbf {\bibinfo {volume} {6}},\ \bibinfo {pages} {376}
  (\bibinfo {year} {2010})}\BibitemShut {NoStop}%
\bibitem [{foo()}]{footnote1}%
  \BibitemOpen
  \href@noop {} {}\bibinfo {note} {Note that the topological Mott insulator
  considered in this paper is different from the proposal of Sri Raghu {\it et
  al.}; for details please see the Supplemental Material.}\BibitemShut {Stop}%
\bibitem [{\citenamefont {Rachel}\ and\ \citenamefont
  {Le~Hur}(2010)}]{rachel10}%
  \BibitemOpen
  \bibfield  {author} {\bibinfo {author} {\bibfnamefont {S.}~\bibnamefont
  {Rachel}}\ and\ \bibinfo {author} {\bibfnamefont {K.}~\bibnamefont
  {Le~Hur}},\ }\href@noop {} {\bibfield  {journal} {\bibinfo  {journal} {Phys.
  Rev. B}\ }\textbf {\bibinfo {volume} {82}},\ \bibinfo {pages} {075106}
  (\bibinfo {year} {2010})}\BibitemShut {NoStop}%
\bibitem [{\citenamefont {Witczak-Krempa}\ \emph {et~al.}(2010)\citenamefont
  {Witczak-Krempa}, \citenamefont {Choy},\ and\ \citenamefont
  {Kim}}]{krempa-10prb165122}%
  \BibitemOpen
  \bibfield  {author} {\bibinfo {author} {\bibfnamefont {W.}~\bibnamefont
  {Witczak-Krempa}}, \bibinfo {author} {\bibfnamefont {T.~P.}\ \bibnamefont
  {Choy}}, \ and\ \bibinfo {author} {\bibfnamefont {Y.~B.}\ \bibnamefont
  {Kim}},\ }\href@noop {} {\bibfield  {journal} {\bibinfo  {journal} {Phys.
  Rev. B}\ }\textbf {\bibinfo {volume} {82}},\ \bibinfo {pages} {165122}
  (\bibinfo {year} {2010})}\BibitemShut {NoStop}%
\bibitem [{\citenamefont {Kargarian}\ \emph {et~al.}(2011)\citenamefont
  {Kargarian}, \citenamefont {Wen},\ and\ \citenamefont {Fiete}}]{fiete11}%
  \BibitemOpen
  \bibfield  {author} {\bibinfo {author} {\bibfnamefont {M.}~\bibnamefont
  {Kargarian}}, \bibinfo {author} {\bibfnamefont {J.}~\bibnamefont {Wen}}, \
  and\ \bibinfo {author} {\bibfnamefont {G.~A.}\ \bibnamefont {Fiete}},\
  }\href@noop {} {\bibfield  {journal} {\bibinfo  {journal} {Phys. Rev. B}\
  }\textbf {\bibinfo {volume} {83}},\ \bibinfo {pages} {165112} (\bibinfo
  {year} {2011})}\BibitemShut {NoStop}%
\bibitem [{\citenamefont {Cho}\ \emph {et~al.}(2012)\citenamefont {Cho},
  \citenamefont {Xu}, \citenamefont {Moore},\ and\ \citenamefont
  {Kim}}]{cho-12njp114030}%
  \BibitemOpen
  \bibfield  {author} {\bibinfo {author} {\bibfnamefont {G.~Y.}\ \bibnamefont
  {Cho}}, \bibinfo {author} {\bibfnamefont {C.}~\bibnamefont {Xu}}, \bibinfo
  {author} {\bibfnamefont {J.~E.}\ \bibnamefont {Moore}}, \ and\ \bibinfo
  {author} {\bibfnamefont {Y.~B.}\ \bibnamefont {Kim}},\ }\href@noop {}
  {\bibfield  {journal} {\bibinfo  {journal} {New. J. Phys.}\ }\textbf
  {\bibinfo {volume} {14}},\ \bibinfo {pages} {115030} (\bibinfo {year}
  {2012})}\BibitemShut {NoStop}%
\bibitem [{\citenamefont {Goldman}\ \emph {et~al.}(2010)\citenamefont
  {Goldman}, \citenamefont {Satija}, \citenamefont {Nikolic}, \citenamefont
  {Bermudez}, \citenamefont {Martin-Delgado}, \citenamefont {Lewenstein},\ and\
  \citenamefont {Spielman}}]{goldman-10prl255302}%
  \BibitemOpen
  \bibfield  {author} {\bibinfo {author} {\bibfnamefont {N.}~\bibnamefont
  {Goldman}}, \bibinfo {author} {\bibfnamefont {I.}~\bibnamefont {Satija}},
  \bibinfo {author} {\bibfnamefont {P.}~\bibnamefont {Nikolic}}, \bibinfo
  {author} {\bibfnamefont {A.}~\bibnamefont {Bermudez}}, \bibinfo {author}
  {\bibfnamefont {M.~A.}\ \bibnamefont {Martin-Delgado}}, \bibinfo {author}
  {\bibfnamefont {M.}~\bibnamefont {Lewenstein}}, \ and\ \bibinfo {author}
  {\bibfnamefont {I.~B.}\ \bibnamefont {Spielman}},\ }\href@noop {} {\bibfield
  {journal} {\bibinfo  {journal} {Phys. Rev. Lett.}\ }\textbf {\bibinfo
  {volume} {105}},\ \bibinfo {pages} {255302} (\bibinfo {year}
  {2010})}\BibitemShut {NoStop}%
\bibitem [{\citenamefont {Cocks}\ \emph {et~al.}(2012)\citenamefont {Cocks},
  \citenamefont {Orth}, \citenamefont {Rachel}, \citenamefont {Buchhold},
  \citenamefont {Le~Hur},\ and\ \citenamefont
  {Hofstetter}}]{cocks-12prl205303}%
  \BibitemOpen
  \bibfield  {author} {\bibinfo {author} {\bibfnamefont {D.}~\bibnamefont
  {Cocks}}, \bibinfo {author} {\bibfnamefont {P.~P.}\ \bibnamefont {Orth}},
  \bibinfo {author} {\bibfnamefont {S.}~\bibnamefont {Rachel}}, \bibinfo
  {author} {\bibfnamefont {M.}~\bibnamefont {Buchhold}}, \bibinfo {author}
  {\bibfnamefont {K.}~\bibnamefont {Le~Hur}}, \ and\ \bibinfo {author}
  {\bibfnamefont {W.}~\bibnamefont {Hofstetter}},\ }\href@noop {} {\bibfield
  {journal} {\bibinfo  {journal} {Phys. Rev. Lett.}\ }\textbf {\bibinfo
  {volume} {109}},\ \bibinfo {pages} {205303} (\bibinfo {year}
  {2012})}\BibitemShut {NoStop}%
\bibitem [{\citenamefont {Orth}\ \emph {et~al.}(2013)\citenamefont {Orth},
  \citenamefont {Cocks}, \citenamefont {Rachel}, \citenamefont {Buchhold},
  \citenamefont {Hur},\ and\ \citenamefont {Hofstetter}}]{orth-13jpb134004}%
  \BibitemOpen
  \bibfield  {author} {\bibinfo {author} {\bibfnamefont {P.~P.}\ \bibnamefont
  {Orth}}, \bibinfo {author} {\bibfnamefont {D.}~\bibnamefont {Cocks}},
  \bibinfo {author} {\bibfnamefont {S.}~\bibnamefont {Rachel}}, \bibinfo
  {author} {\bibfnamefont {M.}~\bibnamefont {Buchhold}}, \bibinfo {author}
  {\bibfnamefont {K.~L.}\ \bibnamefont {Hur}}, \ and\ \bibinfo {author}
  {\bibfnamefont {W.}~\bibnamefont {Hofstetter}},\ }\href@noop {} {\bibfield
  {journal} {\bibinfo  {journal} {J. Phys. B: At. Mol. Opt. Phys.}\ }\textbf
  {\bibinfo {volume} {46}},\ \bibinfo {pages} {134004} (\bibinfo {year}
  {2013})}\BibitemShut {NoStop}%
\bibitem [{\citenamefont {Florens}\ and\ \citenamefont
  {Georges}(2002)}]{florens-02prb165111}%
  \BibitemOpen
  \bibfield  {author} {\bibinfo {author} {\bibfnamefont {S.}~\bibnamefont
  {Florens}}\ and\ \bibinfo {author} {\bibfnamefont {A.}~\bibnamefont
  {Georges}},\ }\href@noop {} {\bibfield  {journal} {\bibinfo  {journal} {Phys.
  Rev. B}\ }\textbf {\bibinfo {volume} {66}},\ \bibinfo {pages} {165111}
  (\bibinfo {year} {2002})}\BibitemShut {NoStop}%
\bibitem [{\citenamefont {Florens}\ and\ \citenamefont
  {Georges}(2004)}]{florens-04prb035114}%
  \BibitemOpen
  \bibfield  {author} {\bibinfo {author} {\bibfnamefont {S.}~\bibnamefont
  {Florens}}\ and\ \bibinfo {author} {\bibfnamefont {A.}~\bibnamefont
  {Georges}},\ }\href@noop {} {\bibfield  {journal} {\bibinfo  {journal} {Phys.
  Rev. B}\ }\textbf {\bibinfo {volume} {70}},\ \bibinfo {pages} {035114}
  (\bibinfo {year} {2004})}\BibitemShut {NoStop}%
\bibitem [{\citenamefont {Zhao}\ and\ \citenamefont
  {Paramekanti}(2007)}]{zhao-07prb195101}%
  \BibitemOpen
  \bibfield  {author} {\bibinfo {author} {\bibfnamefont {E.}~\bibnamefont
  {Zhao}}\ and\ \bibinfo {author} {\bibfnamefont {A.}~\bibnamefont
  {Paramekanti}},\ }\href@noop {} {\bibfield  {journal} {\bibinfo  {journal}
  {Phys. Rev. B}\ }\textbf {\bibinfo {volume} {76}},\ \bibinfo {pages} {195101}
  (\bibinfo {year} {2007})}\BibitemShut {NoStop}%
\bibitem [{\citenamefont {Winkler}(2003)}]{WinklerSOC}%
  \BibitemOpen
  \bibfield  {author} {\bibinfo {author} {\bibfnamefont {R.}~\bibnamefont
  {Winkler}},\ }\href@noop {} {\emph {\bibinfo {title} {Spin-orbit Coupling
  Effects in Two-Dimensional Electron and Hole Systems}}}\ (\bibinfo
  {publisher} {Springer},\ \bibinfo {year} {2003})\BibitemShut {NoStop}%
\bibitem [{\citenamefont {{F}ukui}\ and\ \citenamefont
  {{H}atsugai}(2007)}]{JPSJ.76.053702}%
  \BibitemOpen
  \bibfield  {author} {\bibinfo {author} {\bibfnamefont {T.}~\bibnamefont
  {{F}ukui}}\ and\ \bibinfo {author} {\bibfnamefont {Y.}~\bibnamefont
  {{H}atsugai}},\ }\href@noop {} {\bibfield  {journal} {\bibinfo  {journal}
  {Journal of the Physical Society of Japan}\ }\textbf {\bibinfo {volume}
  {76}},\ \bibinfo {pages} {053702} (\bibinfo {year} {2007})}\BibitemShut
  {NoStop}%
\bibitem [{\citenamefont {Zhang}\ \emph {et~al.}(2012)\citenamefont {Zhang},
  \citenamefont {Hung}, \citenamefont {Tung},\ and\ \citenamefont
  {Chin}}]{Zhang02032012}%
  \BibitemOpen
  \bibfield  {author} {\bibinfo {author} {\bibfnamefont {X.}~\bibnamefont
  {Zhang}}, \bibinfo {author} {\bibfnamefont {C.-L.}\ \bibnamefont {Hung}},
  \bibinfo {author} {\bibfnamefont {S.-K.}\ \bibnamefont {Tung}}, \ and\
  \bibinfo {author} {\bibfnamefont {C.}~\bibnamefont {Chin}},\ }\href@noop {}
  {\bibfield  {journal} {\bibinfo  {journal} {Science}\ }\textbf {\bibinfo
  {volume} {335}},\ \bibinfo {pages} {1070} (\bibinfo {year}
  {2012})}\BibitemShut {NoStop}%
\bibitem [{\citenamefont {Hazlett}\ \emph {et~al.}(2013)\citenamefont
  {Hazlett}, \citenamefont {Ha},\ and\ \citenamefont {Chin}}]{arXiv:1306.4018}%
  \BibitemOpen
  \bibfield  {author} {\bibinfo {author} {\bibfnamefont {E.~L.}\ \bibnamefont
  {Hazlett}}, \bibinfo {author} {\bibfnamefont {L.-C.}\ \bibnamefont {Ha}}, \
  and\ \bibinfo {author} {\bibfnamefont {C.}~\bibnamefont {Chin}},\ }\href@noop
  {} {\bibfield  {journal} {\bibinfo  {journal} {arXiv:1306.4018}\ } (\bibinfo
  {year} {2013})}\BibitemShut {NoStop}%
\end{thebibliography}
%

\pagebreak


\section*{Supplementary material}

# Supplemental Material for
# Dimensional crossover and cold-atom realization of topological Mott insulators

Mathias S. Scheurer,[1] Stephan Rachel,[2] and Peter P. Orth[1]

[1]*Institute for Theory of Condensed Matter, Karlsruhe Institute of Technology (KIT), Karlsruhe, Germany*
[2]*Institute for Theoretical Physics, TU Dresden, 01062 Dresden, Germany*

(Dated: June 28, 2014)

## S1. Ambiguity of the term "Topological Mott Insulator"

In the literature, the term "Topological Mott insulator" is commonly used for two phenomena or states of matter which have nothing in common except that they belong to the very broad field of topological phases.

(i) One was introduced by Sri Raghu and collaborators [1]: they considered electrons on the honeycomb lattice in the presence of (strong) nearest and next-nearest neighbor repulsive interactions. They argued that the second-neighbor repulsions might generate an effective spin-orbit term via spontaneous symmetry breaking leading to quantum anomalous Hall and quantum spin Hall phases. The *topological Mott insulator* of Raghu *et al.* is topologically equivalent to the two-dimensional topological insulator, and the additional word "Mott" only emphasizes that the non-trivial topology is induced by virtue of interactions. Several authors have studied this type of system recently [2–7]. In principle, also the topological Kondo insulator falls into this category [8, 9]. It is important to emphasize that this type of topological Mott insulator is effectively described by a non-interacting topological insulator model.

(ii) The other one was introduced by Dima Pesin and Leon Balents [10] and has been considered in this paper. As motivated in the paper, the *topological Mott insulator* phase of Pesin and Balents is a spin liquid-like state where the charge and spin degrees of freedom have been separated; the charge undergoes a Mott insulator transition while the deconfined spinons still are described by the topological insulator band structure. With other words, the charge degree forms a *Mott* insulator, while the spin degree is a *topological* insulator, both together are, hence, named *topological Mott insulator*. Recently, this state of matter was also investigated by other authors [11–13]. It is important to emphasize that this type of topological Mott insulator merely exists because of the interplay of non-trivial band topology and strong interactions and does *not* have any non-interacting analogue.

## S2. Twisting the Hubbard interaction via permutation of the degrees of freedom

In the following we present more details about how the non-local interaction term

$$H'_U = \frac{U}{2} \sum_{\bm{j},\sigma=\uparrow,\downarrow} (n_{2j_x j_y j_z \sigma} + n_{(2j_x+1)j_y j_z \sigma} - 1)^2, \quad n_{\bm{j}\sigma} = c^\dagger_{\bm{j}\sigma} c_{\bm{j}\sigma}, \qquad (1)$$

considered in this work can be implemented in a cold-atom setup. The procedure we propose is based on interchanging degrees of freedom: As illustrated in Fig. 1, the spin-degree of freedom of the original fermions $c_{\bm{j}\sigma}$ will be encoded by the evenness and oddness of $j_x$ of the new fermions $d_{\bm{j}\mu}$, whereas the parity $(-1)^{j_x}$ of $j_x$ in the original basis is represented by the internal onsite degree of freedom $\mu = \pm$ of the new fermions. By construction, the nonlocal interaction term in Eq. (1), assumes the form of the usual Hubbard onsite interaction in terms of the transformed fermions $d_{\bm{j}\mu}$. Keeping the non-interacting part of the theory $H_{2D} + H_z$ fixed, the full Hamiltonian reads in the new basis

$$H = -\sum_{\bm{j},\bm{j}'} \sum_{\mu,\mu'} \left( d^\dagger_{\bm{j}\mu} \mathcal{A}_{\mu,\mu'}(\bm{j},\bm{j}') d_{\bm{j}'\mu'} + \text{H.c.} \right) + \lambda \sum_{\bm{j}} \sum_{\mu,\mu'} d^\dagger_{\bm{j}\mu} (\sigma_z)_{\mu,\mu'} d_{\bm{j}\mu'} + \frac{U}{2} \sum_{\bm{j}} \left( \sum_\mu d^\dagger_{\bm{j}\mu} d_{\bm{j}\mu} - 1 \right)^2, \qquad (2)$$

where $\mathcal{A}_{\mu,\mu'}(\bm{j},\bm{j}')$ are the transformed hopping amplitudes of $H_{2D} + H_z$ as illustrated in Fig. 1 for all three spatial directions. Redistributing the original spin degree of freedom in the way we propose, the system only contains nearest and next-nearest neighbor hopping elements. It is very important to note that, although the staggered hopping now has the form of a Zeeman term, Eq. (2) is mathematically equivalent to $H_{2D} + H_z + H'_U$ and, consequently, respects time-reversal symmetry. We emphasize that time-reversal is a non-local operation in the new basis interchanging sites with even and odd $j_x$.

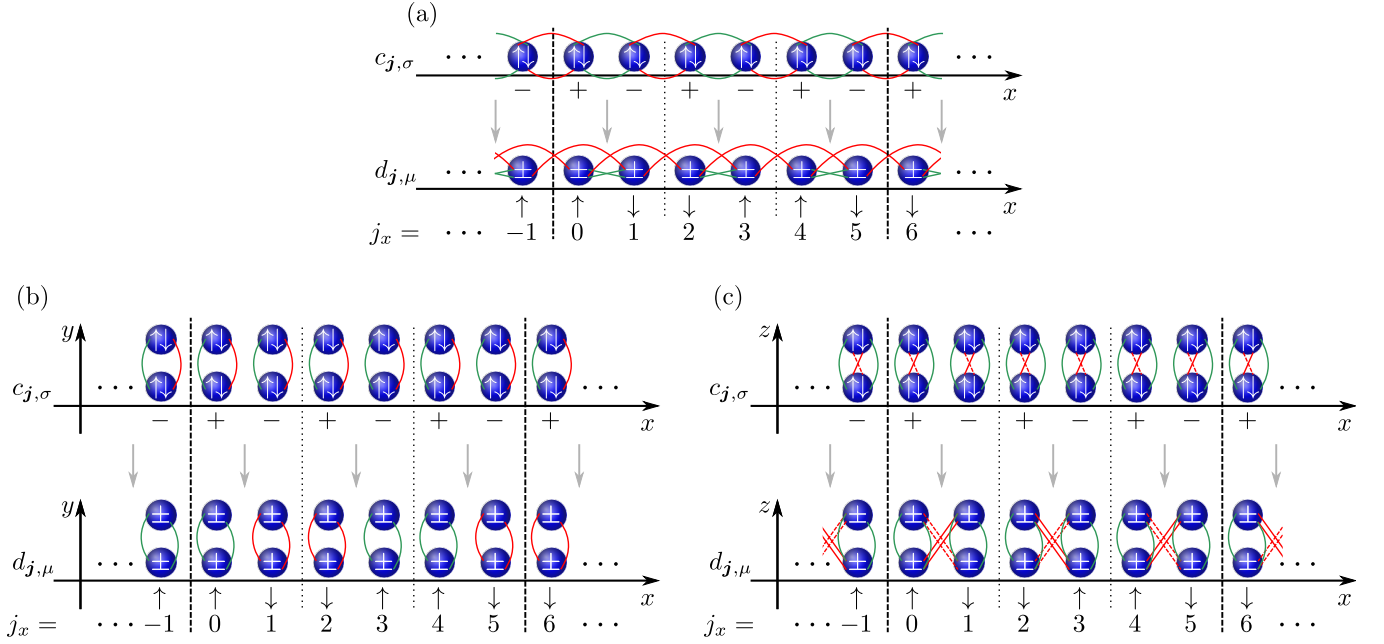

FIG. 1: (Color online) Graphical definition of the transformation to the new lattice and illustration of the required hopping matrix elements along the three different directions. Panels $(a, b, c)$ refer to hopping elements along the $(x, y, z)$ spatial directions. The upper part of the respective panels refers to the hopping elements of the $c_{j\sigma}$ fermions in the Hamiltonian $H_{2D} + H_z$ defined in Eqs. (1) and (2) of the main text. The lower part of the panels shows the required hopping elements of the $d_{j'\mu}$ fermions that have to be implemented in order to obtain the (identical) Hamiltonian $H_{2D} + H_z$ in terms of the $c_{j\sigma}$ fermions if the spin $\sigma$ is encoded spatially (as even and odd sites) and the site parity along $x$, $\mu = \pm$, as the internal hyperfine degree of the freedom.

## S3. Slave-rotor theory

To derive the interacting phase diagrams reported in the main part of the paper, we apply slave-rotor theory [14–16] to the Hamiltonian (2). Within this approach, one introduces phases $\theta_{\boldsymbol{j}}$ conjugate to the total number of fermions on one site of the lattice and auxiliary fermionic operators $f_{\boldsymbol{j}\mu}$ via

$$d_{\boldsymbol{j}\mu} = e^{i\theta_{\boldsymbol{j}}} f_{\boldsymbol{j}\mu}, \qquad d^\dagger_{\boldsymbol{j}\mu} = e^{-i\theta_{\boldsymbol{j}}} f^\dagger_{\boldsymbol{j}\mu}. \tag{3}$$

To restrict the theory to the physical part of the Hilbert space, we have to impose the constraint

$$L_{\boldsymbol{j}} + \sum_{\mu=\pm} f^\dagger_{\boldsymbol{j}\mu} f_{\boldsymbol{j}\mu} = \mathbb{1}, \qquad L_{\boldsymbol{j}} := -i\partial_{\theta_{\boldsymbol{j}}}. \tag{4}$$

This condition will be treated on average and accounted for by introducing Lagrange multipliers $h_{\boldsymbol{j}}$. Passing to a path-integral description, the associated action is given by

$$S_{\text{SR}}[\theta, \bar{f}, f] = \int_0^\beta d\tau \left[ \frac{1}{2U} \sum_{\boldsymbol{j}} (\partial_\tau \theta_{\boldsymbol{j}} - ih_{\boldsymbol{j}})^2 + \sum_{\boldsymbol{j},\mu} \bar{f}_{\boldsymbol{j}\mu} (\partial_\tau - h_{\boldsymbol{j}}) f_{\boldsymbol{j}\mu} \right. \\ \left. - \sum_{\boldsymbol{j},\boldsymbol{j}'} \sum_{\mu,\mu'} \left( \bar{f}_{\boldsymbol{j}\mu} \mathcal{A}_{\mu,\mu'}(\boldsymbol{j},\boldsymbol{j}') e^{i(\theta_{\boldsymbol{j}'} - \theta_{\boldsymbol{j}})} f_{\boldsymbol{j}'\mu'} + \text{G.c.} \right) + \lambda \sum_{\boldsymbol{j}} \bar{f}_{\boldsymbol{j}\mu} (\sigma_z)_{\mu,\mu'} f_{\boldsymbol{j}\mu'} \right], \tag{5}$$

where $f_{\boldsymbol{j}\mu}$, $\bar{f}_{\boldsymbol{j}\mu}$ denote Grassmann variables.

In this work, we use the sigma-model description of Ref. 14, where the phase degrees of freedom are represented by complex bosonic fields

$$X_{\boldsymbol{j}}(\tau) := e^{i\theta_{\boldsymbol{j}}(\tau)} \tag{6}$$

with the nonlinear constraint

$$|X_{\bm{j}}(\tau)|^2 = 1. \tag{7}$$

The latter will be ensured on average via additonal Lagrange multipliers $\rho_{\bm{j}}$. Applying a mean-field approximation to the hopping terms in Eq. (5), we arrive at the bosonic and fermionic actions

$$S_X\left[X_{\bm{j}},X_{\bm{j}}^*\right] = \int_0^\beta d\tau \left[\sum_{\bm{j}} X_{\bm{j}}^*(\tau) \left(\frac{1}{2U}\overleftarrow{\partial_\tau}\overrightarrow{\partial_\tau} + \rho_{\bm{j}}\right) X_{\bm{j}}(\tau) - \sum_{\bm{j},\bm{j}'} \left(X_{\bm{j}}^*(\tau)\mathcal{Q}(\bm{j},\bm{j}')X_{\bm{j}'}(\tau) + \text{c.c.}\right)\right] \tag{8}$$

and

$$S_f\left[\bar{f},f\right] = \int_0^\beta d\tau \left[\sum_{\bm{j},\mu} \bar{f}_{\bm{j}\mu}(\tau)\left(\partial_\tau - \mu\right) f_{\bm{j}\mu}(\tau) - \sum_{\bm{j},\bm{j}'}\sum_{\mu,\mu'} \left(\bar{f}_{\bm{j}\mu}(\tau) Z(\bm{j},\bm{j}') \mathcal{A}_{\mu,\mu'}(\bm{j},\bm{j}') f_{\bm{j}'\mu'}(\tau) + \text{G.c.}\right) \right. \\ \left. + \lambda \sum_{\bm{j},\mu} \bar{f}_{\bm{j}\mu}(\tau) \left(\sigma_z\right)_{\mu,\mu'} f_{\bm{j}\mu'}(\tau)\right], \tag{9}$$

that are coupled via the self-consistency equations

$$\mathcal{Q}(\bm{j},\bm{j}') = \sum_{\mu,\mu'} \langle \bar{f}_{\bm{j}\mu}(\tau) \mathcal{A}_{\mu,\mu'}(\bm{j},\bm{j}') f_{\bm{j}'\mu'}(\tau)\rangle, \tag{10a}$$

$$Z(\bm{j},\bm{j}') = \langle X_{\bm{j}}^*(\tau) X_{\bm{j}'}(\tau)\rangle. \tag{10b}$$

Here we have already taken into account that, at half-filling, the constraint (4) is satisfied on average by choosing $h_{\bm{j}} = 0$. Note that only three of the Lagrange multipliers $\rho_{\bm{j}}$, e.g. $\rho_{0,0,0}$, $\rho_{2,0,0}$ and $\rho_{4,0,0}$, are independent which is due to the combination of translation symmetry and time-reversal invariance.

For any given $U$ one has to solve the set of equations (10) and find the correct Lagrange multipliers $\rho_{0,0,0}$, $\rho_{2,0,0}$, $\rho_{4,0,0}$ yielding the renormalization factors $Z(\bm{j},\bm{j}')$ of the auxiliary fermions $f_{\bm{j}\mu}$. The band structure resulting from Eq. (9) can again be classified exactly as for non-interacting fermions leading to the variety of different correlated topologically trivial and non-trivial phases discussed in the main part of the paper. The transition into the Mott phase, where the local number degree of freedom is frozen out, occurs when the gap of the bosons in Eq. (8) closes. This point marks the transition from a ferromagnetic to a paramagnetic state of the rotors $X_{\bm{j}}$. Even in the paramagnetic state there exist short range correlations between the rotors which implies that the renormalization factors $Z(\bm{j},\bm{j}')$ are non-zero. A topologically non-trivial spinon bandstructure while having paramagnetic rotors defines the weak and strong topological Mott insulator phases.

[1]  S. Raghu, X.-L. Qi, C. Honerkamp, and S.-C. Zhang, Phys. Rev. Lett. **100**, 156401 (2008).
[2]  C. Weeks and M. Franz, Phys. Rev. B **81**, 085105 (2010).
[3]  S. Uebelacker and C. Honerkamp, Phys. Rev. B **84**, 205122 (2011).
[4]  A. Dauphin, M. Müller, and M. A. Martin-Delgado, Phys. Rev. A **86**, 053618 (2012).
[5]  M. Daghofer and M. Hohenadler, Phys. Rev. B **89**, 035103 (2014).
[6]  N. A. Garcia-Marinez, A. G. Grushin, T. Neupert, B. Valenzuela, and E. V. Castro, Phys. Rev. B **88**, 245123 (2013).
[7]  T. Duric, N. Chancellor, and I. F. Herbut, Phys. Rev. B **89**, 165123 (2014).
[8]  M. Dzero, K. Sun, V. Galitski, and P. Coleman, Phys. Rev. Lett. **104**, 106408 (2010).
[9]  M. Neupane and *et al.*, Nat. Comm. **4**, 2991 (2013).
[10] D. A. Pesin and L. Balents, Nature Phys. **6**, 376 (2010).
[11] W. Witczak-Krempa, T. P. Choy, and Y. B. Kim, Phys. Rev. B **82**, 165122 (2010).
[12] S. Rachel and K. Le Hur, Phys. Rev. B **82**, 075106 (2010).
[13] M. Kargarian, J. Wen, and G. A. Fiete, Phys. Rev. B **83**, 165112 (2011).
[14] S. Florens and A. Georges, Phys. Rev. B **66**, 165111 (2002).
[15] S. Florens and A. Georges, Phys. Rev. B **70**, 035114 (2004).
[16] E. Zhao and A. Paramekanti, Phys. Rev. B **76**, 195101 (2007).



\end{document}